\documentclass[12pt]{iopart}

\usepackage[numbers, compress]{natbib}

\usepackage[utf8]{inputenc} % allow utf-8 input
\usepackage[T1]{fontenc}    % use 8-bit T1 fonts
\usepackage[
    colorlinks=true,
    allcolors=C1]{hyperref}
\usepackage{url}            % simple URL typesetting

\usepackage{breakurl} % allow autmatic break in long bibliography URLs
\usepackage{booktabs}       % professional-quality tables
\usepackage{amsfonts}       % blackboard math symbols
\usepackage{nicefrac}       % compact symbols for 1/2, etc.
\usepackage{microtype}      % microtypography
\usepackage{graphicx}       % figures

\usepackage{amssymb}
\usepackage{bm}
\usepackage{braket}

\newcommand{\infdivx}[2]{(#1 \; \| \; #2)}
\newcommand{\kld}{D_\mathrm{KL}\infdivx}

\usepackage{placeins}
\usepackage{ragged2e}

\usepackage{subcaption}
\usepackage[symbol]{footmisc}

\usepackage[table]{xcolor}
\usepackage{multicol}
\usepackage{tabularx}
\usepackage{siunitx}

\definecolor{C0}{HTML}{ab3e2e}  % red
\definecolor{C1}{HTML}{2749e8}  % blue
\definecolor{acclightblue}{RGB}{146, 218, 221}
\definecolor{accblue}{RGB}{63, 144, 218}

\usepackage{setspace}

\begin{document}

\title[Flow annealed importance sampling bootstrap meets differentiable particle physics]{Flow annealed importance sampling bootstrap meets differentiable particle physics}

\author{A Kofler$^{1,2,3}$, V Stimper$^{1,4,5}$, M Mikhasenko$^{6,7}$ and M Kagan$^{8}$ and L Heinrich$^3$}

\address{$^1$Max Planck Institute for Intelligent Systems Tübingen, Max-Planck-Ring 4, 72076 Tübingen, Germany}
\address{$^2$Max Planck Institute for Gravitational Physics (Albert Einstein Institute), Potsdam Science Park, Am Mühlenberg 1, 14476 Potsdam, Germany}
\address{$^3$Technical University of Munich, TUM School of Natural Sciences, Physics Department, James-Franck-Str. 1, 85747 Garching, Germany}
\address{$^4$Isomorphic Labs, 280 Bishopsgate, London, EC2M 4RB, United Kingdom}
\address{$^5$University of Cambridge, Department of Engineering, Trumpington Street, Cambridge, CB2 1PZ, UK}
\address{$^6$University of Bochum, Universitätsstraße 150,
44801 Bochum, Germany}
\address{$^7$Excellence Cluster ORIGINS, Boltzmannstr. 2, D-85748 Garching, Germany}
\address{$^8$SLAC National Accellerator Laboratory, 2575 Sand Hill Rd, Menlo Park, CA 94025-7015, USA}

\ead{l.heinrich@tum.de}
\vspace{10pt}
\begin{indented}
\item[]Submitted for review: February 2025, Accepted for publication: May 2025
\end{indented}

\begin{abstract}
    High-energy physics requires the generation of large numbers of simulated data samples from complex but analytically tractable distributions called matrix elements. 
    Surrogate models, such as normalizing flows, are gaining popularity for this task due to their computational efficiency.
    We adopt an approach based on Flow Annealed importance sampling Bootstrap~(FAB) that evaluates the differentiable target density during training and helps avoid the costly generation of training data in advance.
    We show that FAB reaches higher sampling efficiency with fewer target evaluations in high dimensions in comparison to other methods. 
\end{abstract}

\vspace{2pc}
\noindent{\it Keywords}: Normalizing Flows, Particle Physics, Differentiable Programming, Annealed Importance Sampling, Generative Modeling

%\maketitle
%\ioptwocol % TODO: remove after testing figures and tables
%
\vfill
\begin{spacing}{0.75}
{
\noindent\rule{\linewidth}{0.4pt}
\scriptsize
This is the Accepted Manuscript version of an article accepted for publication in \textit{Machine Learning: Science and Technology}. IOP Publishing Ltd is not responsible for any errors or omissions in this version of the manuscript or any version derived from it. This Accepted Manuscript is published under a CC~BY licence. The Version of Record is available online at \href{https://doi.org/10.1088/2632-2153/addbc1}{10.1088/2632-2153/addbc1}.
}
\end{spacing}
% In this work, we (1)~adopt FAB to HEP, (2)~benchmark it against other training methods, and (3)~provide a detailed performance comparison based on the number of target evaluations during training.

\section{Introduction}

In the advent of the high-luminosity phase at the Large Hadron Collider (LHC), significant speed-ups in the simulation software are required to analyze the increasing amount of data~\citep{aberle:2020, hl-lhc-roadmap_CMS:2022, hl-lhc-roadmap_ATLAS:2022}. 
The simulated data are compared to measured data from particle collisions to understand the underlying fundamental physics processes in more detail.
One important step in the LHC simulation chain is the generation of samples (``events'') based on \emph{matrix elements} (MEs).
MEs describe the dynamical information contained in particle interactions.
The ME $\mathcal{M} = \Braket{p_1,..., p_n| \mathcal{M}| p_a, p_b}$ quantifies the transition from an initial state with momenta $p_a$ and $p_b$ to a final state with $n$ outgoing particles described by their momenta $p_1, p_2,..., p_n$.
Using quantum field theory, MEs can be constructed by summing the contributions of all possible Feynman diagrams $\mathcal{M} = \mathcal{M}_1 + \mathcal{M}_2 + \mathcal{M}_3 + ...$ (also called ``channels''). Individual MEs can be calculated using the Feynman rules~\citep{griffith:1987}, which can become computationally expensive to evaluate when higher-order terms are included.
The dimensionality of the matrix element depends on the number of outgoing particles $n$, each having three spatial degrees of freedom. Taking mass and momentum constraints into account, the dimensionality of MEs amounts to~$3n -4$.
From a machine learning perspective, one can interpret MEs as unnormalized distributions $p(x)$ over the outgoing 4-momenta $\{p_1,..., p_n\}$ which we will denote as $x$ in the following to simplify notation.
Via the Feynman rules, MEs can be evaluated analytically; however, sampling from them is hard since they can be high-dimensional for large $n$ and multi-modal due to contributions from different channels. 
Additionally, they are defined on limited support originating from mass and momentum constraints and can exhibit divergences.
It is possible to simplify the complicated multi-modal structure of MEs by decomposing the ME into its dominant channels, an approach referred to as multi-channeling~\citep{kleiss:1994}.

Standard sampling algorithms such as \texttt{MadGraph}~\citep{alwall:2011, alwall:2014}, \texttt{SHERPA}~\citep{sherpa:2019}, and \texttt{PYTHIA}~\citep{bierlich:2022_pythia, bierlich:2020_pythia_code} rely on multi-channeling to reduce the complexity of the ME and employ adaptive Monte Carlo methods similar to \texttt{VEGAS}~\citep{lepage:1978} to approximate the distribution of the individual channels.
More recently, machine learning-based surrogate models like normalizing flows have shown to improve the efficiency of the sampling process compared to standard methods like~\texttt{VEGAS} and multiple approaches have been proposed over the last years.
In general, normalizing flows can be trained using divergence measures as loss functions that quantify the difference between the target and the flow distribution: 
Information about the target is either included by training with samples from the target or by evaluating the target distribution analytically.
When training normalizing flows with the maximum-likelihood loss based on target samples~\citep{stienen:2021}, we have to rely on the costly generation of a large training data set. Therefore, approaches that evaluate the distribution of interest directly during training with samples from the flow, have been the focus in event generation:
Most papers employ neural importance sampling~(NIS)~\citep{mueller:2019} which has been adopted to phase space sampling and combined with multi-channeling in the works of~\citep{bothmann:2020, gao:2020_iflow, gao:2020_evgen_nfs}.
Building on these developments, the efficiency of this approach can be improved by introducing a replay buffer and a \texttt{VEGAS} based initialization, as well as adapting the coupling between dimensions in the normalizing flow~\citep{heimel:2023, heimel:2024}.
NIS without multi-channeling has been employed in the work of~\citep{pina-otey:2020} where they extend the target to include a background density and develop a specific training scheme to boost the efficiency at the beginning of training. 
To make previous results accessible to non-specialist users, a dedicated library has been developed recently~\citep{deutschmann:2024}.

In this work, we take into account that gradients need to be backpropagated through the ME distribution to update the flow parameters when evaluating the target with samples from the flow.
Therefore, this training mode usually requires differentiable MEs, which have only recently been proposed~\citep{heinrich:2022} and employed~\citep{heimel:2024_diffable} for normalizing flow training.
A similar method, called Flow Annealed importance sampling Bootstrap~(FAB)~\citep{midgley:2023}, also relies on the evaluation of a differentiable target density and has been developed to obtain samples from Boltzmann distributions of molecules. 
FAB uses Annealed Importance Sampling~(AIS) with Hamiltonian Monte Carlo~(HMC) transition steps to improve the quality of the normalizing flow samples towards the distribution of interest. 
The resulting AIS samples and their weights are used to train the normalizing flow.
Running HMC requires a differentiable target distribution, and we are the first to perform HMC-based updates on MEs.
To compare the general performance of different training methods, we do not include domain-specific physics information via multi-channeling.

\paragraph{Contributions.}
In this work, (1)~we adopt FAB~\citep{midgley:2023} to event generation in the field of high-energy physics~(HEP).
(2)~We compare FAB with an alternative density evaluation-based method using the reverse Kullback-Leibler Divergence (rKLD)~\citep{papamakarios:2021} as a loss function, as well as with sample-based maximum-likelihood training with the forward Kullback-Leibler Divergence~(fKLD) loss. 
Finally, (3)~we provide a detailed performance comparison between the methods based on the number of ME evaluations, as this step can be expensive.

\section{Method} 
\label{section:method}

A normalizing flow~\citep{papamakarios:2021, tabak:2010, rezende:2015} is a density estimator that consists of a series of learnable, invertible transformations that construct a differentiable bijection between a simple, chosen base distribution and an expressive flow distribution~$q_\theta(x)$.
To generate samples $x \sim q_\theta(x)$, samples are obtained from the base distribution and passed through the subsequent transformations. 
The density~$q_\theta(x)$ can be evaluated for a given data point~$x$ by passing it through the inverse chain of transformations and evaluating the density of the base distribution.

\subsection{Training}
To optimize the flow parameters~$\theta$ such that $q_\theta(x)$ matches the (unnormalized) target distribution~$p(x)$ more closely, different forms of the Kullback-Leibler~(KL) divergence are utilized, resulting in two main approaches:
Normalizing flows can either be trained with available samples from the target distribution~$x \sim p(x)$ (fKLD), or by evaluating the density~$p(x)$ with samples from the flow~$x \sim q_\theta(x)$ (rKLD and FAB).

%\vspace{-0.15cm}
\paragraph{Forward KL Divergence (fKLD).}
If samples~$x \sim p(x)$ are cheaply available, the fKLD serves as a loss function: 
\begin{equation*}
    \kld{p}{q_\theta} = \mathbb{E}_{x \sim p(x)}\left[\log \frac{p(x)} {q_\theta(x)}\right]~.
\end{equation*}
It can be simplified to the negative log-likelihood loss
\begin{equation*}
    \mathcal{L}_\mathrm{fKLD} = - \mathbb{E}_{x \sim p(x)}\left[\log q_\theta(x)\right] = - \frac{1}{N} \sum_{i=1}^N \log q_\theta(x_i)
\end{equation*}
over $N$ data points where we drop terms independent of~$\theta$. 
It can be shown that this loss function results in a density~$q_\theta$ with mass-covering properties~\citep{midgley:2023}. 
The flow distribution~$q_\theta(x)$ is evaluated with samples from~$p(x)$ and the gradient computation is straightforward:
\mbox{$\nabla_\theta \mathcal{L}_\mathrm{fKLD} = - \frac{1}{N} \sum_{i=1}^N \nabla_\theta \log q_\theta(x_i)~.
$}

\begin{figure*}[!ht]
    \centering
    \begin{subcaptionblock}{.5\linewidth}
        \centering
        \includegraphics{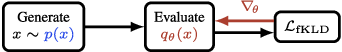}
        \caption{Training with fKLD}
        \label{fig:visualization_fkld}
    \end{subcaptionblock}%
    \begin{subcaptionblock}{.5\linewidth}
        \centering
        \includegraphics{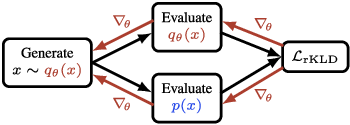}
        \caption{Training with rKLD}
        \label{fig:visualization_rkld}
    \end{subcaptionblock}\\[6pt]
    \begin{subcaptionblock}{\textwidth}
        \centering
        \includegraphics{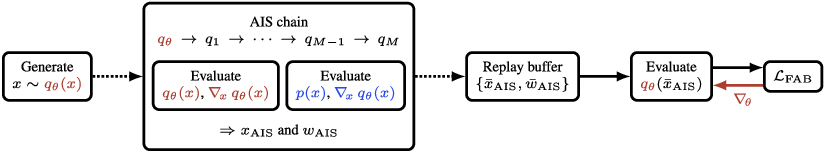}
        \caption{Training with FAB}
        \label{fig:visualization_fab}
    \end{subcaptionblock}\\
    \caption{
        Visualization of compared methods. 
        The black arrows illustrate the forward pass, while the red arrows mark the backpropagation of the gradient to update the flow parameters~$\theta$. 
        For rKLD, the gradient propagation to the normalizing flow~$q_\theta(x)$~(\textcolor{C0}{red}) requires a differentiable target distribution~$p(x)$~(\textcolor{C1}{blue}). 
        For FAB, the HMC updates compute the gradient of the distribution of interest, while the backpropagation of the gradient is stopped for the AIS samples~$\bar{x}_\mathrm{AIS}$ and weights~$\bar{w}_\mathrm{AIS}$. 
        The dotted lines indicate that these steps do not occur when a gradient update is performed repeatedly with samples from the replay buffer. 
    }
    \label{fig:visualization_methods}
\end{figure*}

\paragraph{Reverse KL Divergence (rKLD).}
Another way of quantifying the difference between two distributions is via the rKLD, 
\begin{equation*}
    \kld{q_\theta}{p} = \mathbb{E}_{x\sim q_\theta(x)}\left[\log \frac{q_\theta(x)}{p(x)}\right]~.
\end{equation*}
Compared to fKLD, we sample from the normalizing flow and evaluate the flow density~$q_\theta(x)$ as well as the target distribution~$p(x)$ with the obtained samples.
As a result, $p(x)$ has to be analytically available which is the case for MEs.
This loss function has mode-seeking properties, meaning that it is not guaranteed that the optimized distribution $q_\theta(x)$ covers all modes of $p(x)$~\citep{midgley:2023}.
To obtain a gradient for this loss function
\begin{equation*}
    \nabla_\theta \mathcal{L}_\mathrm{rKLD} = \nabla_\theta \mathbb{E}_{x \sim q_\theta(x)}\left[\log \frac{q_\theta(x)}{p(x)}\right]~,
\end{equation*}
the gradient has to be propagated through the evaluation of the target distribution~$p(x)$, since the samples~$x \sim q_\theta(x)$ used for the estimation of the expectation value depend on the parameters~$\theta$ themselves. 
This is visualized in figure~\ref{fig:visualization_rkld} for illustration.
As a result, $p(x)$ has to be differentiable\footnote{One can prevent differentiating through the target distribution by mapping the flow samples (with stopped gradient) back through the flow in the reverse direction and evaluating the loss function in latent space \cite{heimel:2023}.
However, this approach requires two subsequent applications of the flow transformations and is more costly.}.

%\vspace{-0.15cm}
\paragraph{FAB.}
Both versions of the KL~divergence are special cases of the $\alpha$-divergence~\citep{minka:2005}
\begin{equation}
    D_{\alpha} \infdivx{p}{q_\theta} = - \frac{1}{\alpha(1-\alpha)} \int p(x)^\alpha ~ q_\theta(x)^{1 - \alpha} ~\mathrm{d}x~,
\end{equation}
where fKLD corresponds to~$\alpha \rightarrow 1$ and rKLD to~$\alpha \rightarrow 0$~\citep{minka:2005}.
In FAB~\citep{midgley:2023}, \mbox{$D_{\alpha=2}$} is chosen as a loss function since it minimizes the variance of the importance weights~\mbox{$w = p(x) / q_\theta(x)$}---a~desirable property of a well-performing density estimator $q_\theta(x)$---and the resulting distribution~$q_\theta(x)$ has mass covering properties with respect to~$p(x)$~\citep{minka:2005}.
Since samples from the flow might poorly fit the target distribution at the beginning of training, AIS~\citep{neal:2001} is employed to pass the samples through a chain of intermediate distributions~$q_1,..., q_{M-1}$ with HMC as a transition operator. 
The intermediate distributions $q_i$ are chosen to interpolate between the flow distribution $q_0 = q_\theta$ and the AIS target $q_M = p^2 / q_\theta$. 
For all HMC steps between two intermediate AIS distributions, the flow as well as the target distribution have to be evaluated and the gradient of both distributions is required to perform momentum updates. As a result, we require a differentiable target distribution for FAB with HMC.
Importance weights~$w_\mathrm{AIS}$ can be computed for the AIS samples $x_\mathrm{AIS}$ which allow evaluating the surrogate loss function
\begin{equation*}
    \mathcal{S}(\theta) = - \mathbb{E}_{\mathrm{AIS}} \left[\bar{w}_\mathrm{AIS} \log q_\theta(\bar{x}_\mathrm{AIS})\right]~,
\end{equation*}
where the bar over variables indicates that the gradient is not propagated through the AIS sequence in the backward pass.
To reduce the cost of evaluating the target distribution and its gradient for each intermediate distribution and each HMC step in between, the AIS samples can be stored in a replay buffer. 
Pairs~$\{\bar{x}_\mathrm{AIS}, \bar{w}_\mathrm{AIS}\}$ are sampled from the buffer based on the importance weights to perform multiple gradient updates per iteration.
Figure~\ref{fig:visualization_fab} shows a schematic illustration of FAB; for further details, see~\citep{midgley:2023}.

\subsection{Differentiable Matrix Elements}
\begin{figure*}[ht]
    \centering
    \includegraphics[width=\textwidth]{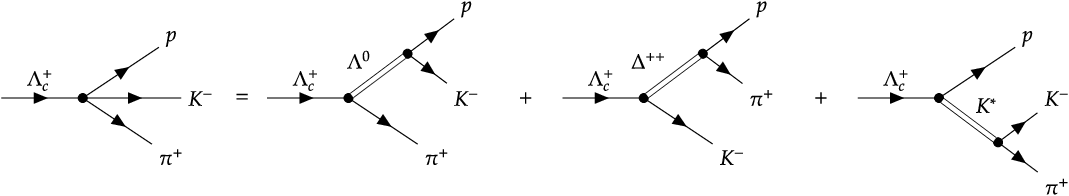}
    \caption{The $\Lambda_c^+ \rightarrow pK^- \pi^+$ decay (left-most diagram) consists of  three decay channels characterized by their resonances $\Lambda_0$ (center-left), $\Delta^{++}$ (center-right), and $K^*$ (right), each visualized in Feynman-like diagrams.}
    \label{fig:lambdac_decay}
\end{figure*}

The training approaches explored and compared in this work are not limited to one area of event generation in particle physics. 
They only require a differentiable implementation of the matrix element, since training with rKLD and FAB requires gradients of the target distribution. 
Recent developments in HEP provide us with differentiable implementations of complex amplitudes~\citep{heinrich:2022, compwa:2023, heimel:2024_diffable}.
We choose two examples of high interest for the particle physics community, one from flavor physics and one from collider physics:

\paragraph{$\Lambda_c^+ \rightarrow pK^-\pi^+$.} 
In flavor physics, the resonances of baryons are studied to search for new states beyond the constituent-quark model~\citep{adoplh:2017, kaspar:2022}. 
One recently analyzed example is the $\Lambda_c^+$ baryon and its decay into a proton, a kaon, and a pion~\citep{marangotto:2020, aaij:2023, LHCb:2023crj}. 
This decay is particularly interesting since its resonance structure allows searches of \textit{CP}~symmetry violations~\citep{aaij:2023} and new physics~\citep{penalva:2019, hu:2021}.
When the $\Lambda_c^+$ baryon decays, it can do so via different decay channels, resulting in the $K^- \pi^+$, $p K^-$, and $p \pi^+$ systems. These decays are shown in figure~\ref{fig:lambdac_decay} as Feynman-like diagrams.
Each of these systems decays via an intermediate particle that can have resonances which are shown in a Dalitz plot in figure~\ref{fig:lambdac_visualization}.
A Dalitz plot is a physics-specific visualization of 2D amplitudes where the axes are chosen such that the histograms can be interpreted as an unnormalized density~\citep{byckling:1973}. 
The resonances appear in a Dalitz plot as vertical, horizontal, and diagonal structures.
In figure~\ref{fig:lambdac_visualization}, the vertical bands with two prominent lines correspond to the decay of $\Lambda \rightarrow pK^-$ resulting from the $\Lambda(1520)$ and the $\Lambda(1670)$~resonances.
The significant horizontal band corresponds to the $K^*(892)$ meson based on $K^* \rightarrow K^- \pi^+$. 
Decays in $\Delta^{++} \rightarrow p \pi^+$ are visible on the diagonal of the Dalitz plot where the $\Delta^{++}(1232)$~resonance is notable.
An example for interference effects among resonances of different decay channels is the horizontal $K^*(892)$~band that gets shifted when crossing the vertical line of the $\Lambda(1670)$~resonance. 
Destructive interference is visible at the upper corner of the Dalitz plot resulting from reciprocal influences of $\Lambda(1520)$ and higher-mass $K^*$~resonances~\cite{aaij:2023}.
The complex resonance structure of this decay is ideally suited for our study as it provides a low-dimensional, but challenging setting to compare the performance of different normalizing flow training methods. 
Additionally, the $\Lambda_c^+$~amplitude model is implemented in a differentiable way and the code is publicly available via the \texttt{ComPWA} package~\citep{compwa:2023}.

For the implementation of the $\Lambda_c^+$~decays, \texttt{ComPWA} utilizes the \texttt{SymPy} engine~\cite{sympy:2017} to formulate symbolic amplitude models whose compute graph can be transformed to \texttt{JAX}~\cite{jax:2018} for efficient gradient computation. 
Since \texttt{ComPWA} was developed as a general purpose tool for analytically formulating amplitudes, researchers can implement new amplitude models.
The specific shape of the Dalitz plot boundary originates from mass and momentum constraints.
It would be suboptimal to train a normalizing flow directly on the Dalitz plot representation because the it could assign non-zero probability density outside the physically-plausible regions.
To prevent this, we map the support of the ME to the unit interval in each dimension via a differentiable and invertible transformation which is explained in~\ref{sec:trafo_DP_to_unit_hypercube}.
Similar transformations are applied in standard samplers~\citep{sherpa:2019, madgraph:2014} and we employ rational-quadratic spline flows because they are explicitly designed for distributions with limited support. 
It is not possible to map the distribution to infinite support since kinematic divergences can occur at the phase space boundary~\citep{griffith:1987}, which would result in non-zero probability mass at infinity.

\paragraph{$e^+e^- \rightarrow t\bar{t}, t\rightarrow W^+ b, \bar{t} \rightarrow W^- \bar{b}$.} 
In collider physics, the $t\bar{t}$ reaction producing a $W$~boson and a bottom quark is repeatedly investigated due to the coupling of non-standard model processes to heavy elementary particles like the top quark~\cite{d0:2010ocb, CMS:2011woj, aad:2022jbj, aad:2023}.
We employ $t \bar{t}$ production with unstable final states containing $W$ bosons from an electron positron collider as a challenging, higher dimensional example. 
We selected an $e^+ e^-$ process since ML methods in event generation are relevant for future circular or linear electron-positron colliders~\citep{andre:2025, balazs:2025}.
The eight-dimensional $t\bar{t}$ matrix element can be generated with \texttt{MadJAX}~\citep{heinrich:2022} which provides a differentiable implementation of the scattering amplitudes of \texttt{MadGraph}~\citep{alwall:2011, alwall:2014}.
With \texttt{MadJAX}, we can generate any leading-order amplitudes with spin~$0, 1/2$ or~$1$~particles and the code can be extended to higher spin MEs.
It is also possible to generate MEs for proton-proton collisions by including a differentiable implementation of the parton density~\citep{carrazza:2021, heimel:2024_diffable} which we leave for future work.
To map the irrgeular phase space boundaries to the unit hypercube, a differentiable non-linear transformation based on the \texttt{RAMBO} algorithm is applied to the \texttt{MadJAX} ME~\citep{plaetzer:2013}. 
The eight dimensions characterizing the final state of the selected ME can be interpreted as two rescaled, intermediate masses $\widetilde{M}_1$ and $\widetilde{M}_2$, as well as two angular variables $\cos \widetilde{\theta}_k$ and $\widetilde{\phi}_k$ with $k \in [1,2,3]$.

To summarize, both selected examples are based on libraries which allow users to generate differentiable implementations of arbitrary MEs.
This means that the compared approaches of training normalizing flows can easily be applied to different MEs.

\section{Experimental Setup}
\paragraph{Training Settings.}
For each of the aforementioned MEs and training methods, we train three models with different random seeds and average their results to improve reliability. 
For details about the flow hyperparameters and train settings, we refer to
\ref{sec:hyperparameters} for details.

\paragraph{Baseline.}
We compare our results with a physics-agnostic integral estimation method called \texttt{VEGAS+}~\citep{lepage:1978, lepage:2021}. 
This grid-based optimization method divides the support into a regular, rectangular grid and estimates the integral contribution of each subspace. 
With this information, the grid is iteratively updated to focus on regions with large contributions and the value for the integral estimate is calculated as a weighted average over multiple runs.
Through a combination of stratified and importance sampling, \texttt{VEGAS+} is fast and efficient and can account for correlations between dimensions.
We describe the \texttt{VEGAS+} specific hyperparameters and tuning procedure in detail in~\ref{sec:hyperparameters}.
It is not possible to use \texttt{MadGraph} as a baseline for both reactions since it is tailored to collider physics and the complicated helicity structure of the $\Lambda_c^+$~decay is not implemented.

\paragraph{Performance Metrics.}
We first provide qualitative performance comparisons for both MEs by showing histograms and a corner plot of samples obtained from \texttt{VEGAS+} as well as the normalizing flows.
For the $\Lambda_c^+$ ME, we compare the results to samples obtained by rejection sampling with a uniform proposal distribution.
For the $t\bar{t}$ reaction, we include the samples obtained from~\texttt{MadGraph} as a reference.
Both sets of samples correspond to the training data sets employed for fKLD training.
Secondly, we evaluate the quantitative performance of the normalizing flows based on three metrics: 
(1)~the fKLD evaluated on a test data set, (2)~the importance sampling efficiency calculated from the importance weights, and (3)~the integral estimate of $p(x)$.
The fKLD, introduced in Section~\ref{section:method}, quantifies the difference between the learned distribution $q_\theta(x)$ and the true target distribution~$p(x)$ by evaluating the normalizing flow with the samples~$x \sim p(x)$. 
If both distributions match almost everywhere, the KL divergence is zero.

The importance sampling efficiency~\citep{martino:2017, elvira:2022} can be computed with samples from the normalizing flow $x_i\sim q_\theta(x_i)$ and their importance weights $w_i = p(x_i) / q_\theta(x_i)$ as 
\begin{equation}
    \epsilon = \frac{1}{N}\left[ \sum_{i=1}^N w_i \right]^2 \left[ \sum_{i=1}^N w_i^2 \right]^{-1}\,. 
    \label{eq:sampling_efficiency}
\end{equation}
We elaborate on its derivation and similarities to the unweighting efficiency---which is commonly employed in HEP---in  \ref{sec:efficiencies}.
The integral estimate~$\bar{I}$ of the target distribution $p(x)$ is defined as 
\begin{equation}
    \bar{I} = \int p(x) ~\mathrm{d}x = \int q_\theta(x) \frac{p(x)}{ q_\theta(x)} ~\mathrm{d}x \approx \frac{1}{N} \sum_{i=1}^N w_i~.
\end{equation}
For the 2D~example, the sampling efficiency~$\epsilon$ and the integral estimate~$\bar{I}$ are calculated for $10^4$~flow samples, while $10^6$~flow samples are used for the 8D~ME.
We compare our results for~$\bar{I}$ with \texttt{VEGAS+}.
The performance metrics are summarized in table~\ref{table:performance_metrics} for both MEs. To make rKLD and FAB comparable, we report results based on training runs with the same number of target evaluations.
Since the target distribution can be costly to evaluate, we additionally show the trend for the importance sampling efficiency~$\epsilon$ as a function of the number of target evaluations. 
We do not include fKLD training in this consideration because it relies on a pre-computed training data set on which the normalizing flow can potentially be trained for an arbitrary number of epochs.
For rKLD, the ME is calculated once for each batch of flow samples, while the target density and its gradient need to be evaluated for every HMC step between intermediate AIS distributions for FAB. 
To reduce the number of target evaluations, FAB allows multiple gradient updates per iteration with samples from the replay buffer. 
However, depending on the buffer size, more target evaluations might be required at the beginning of training to fill up the buffer.

\section{Results and Discussion}

\begin{figure*}[ht]
    \centering
    \includegraphics[width=\textwidth]{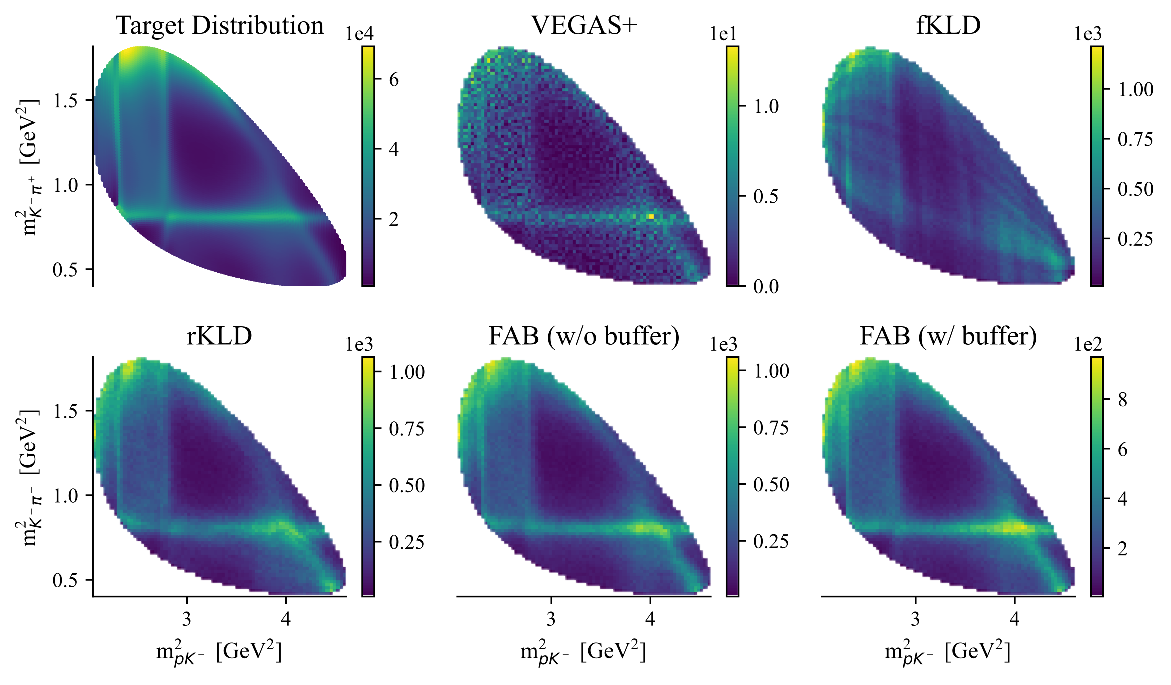}
    \caption{Comparison of the target density for the $\Lambda_c^+ \rightarrow pK^- \pi^+$ matrix element with histograms based on samples from \texttt{VEGAS+} and the best normalizing flow for each method.}
    \label{fig:lambdac_visualization}
\end{figure*}

\begin{figure*}[ht]
    \centering
    \includegraphics[width=0.9\textwidth]{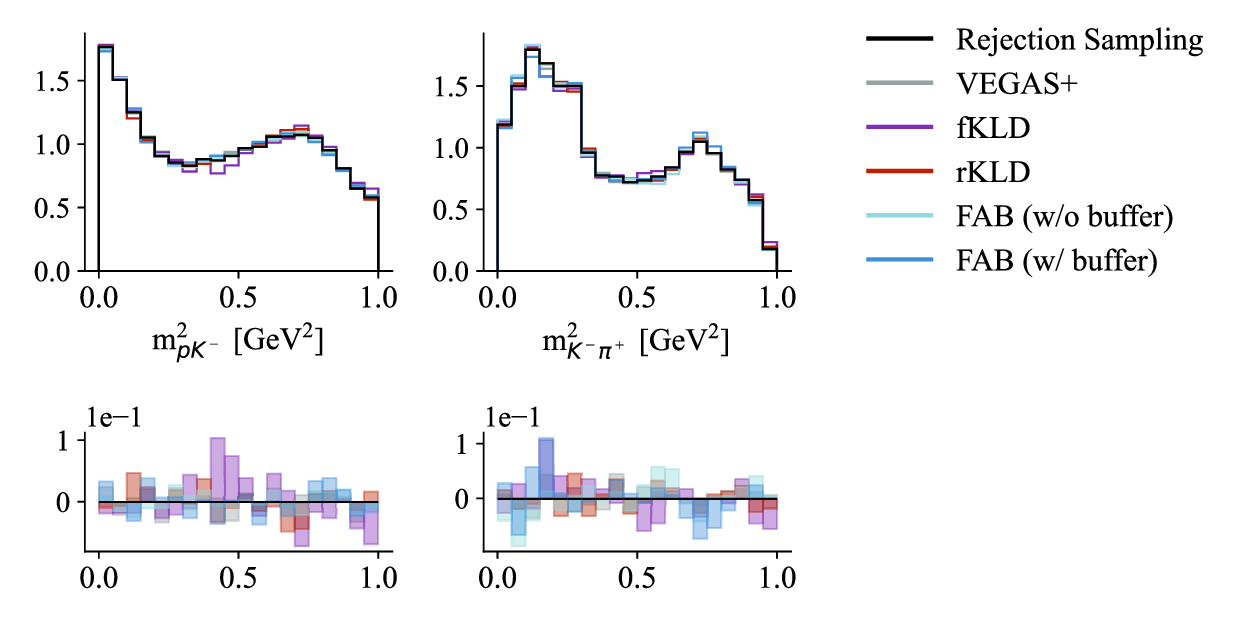}
    \caption{Normalized marginal histograms of the invariant masses for the $\Lambda_c^+ \rightarrow pK^- \pi^+$ matrix element based on samples from rejection sampling, \texttt{VEGAS+}, and the best normalizing flow for each method. The rejection sampling result serves as a baseline to visualize the deviation.}
    \label{fig:lambdac_marginal_hist}
\end{figure*}

\paragraph{$\Lambda_c^+ \rightarrow pK^-\pi^+$.}
To illustrate the complexity of the two-dimensional ME, we show the target density evaluated on a grid as a Dalitz plot in figure~\ref{fig:lambdac_visualization}. 
We provide histograms of $10^6$ unweighted samples obtained from the best performing normalizing flow for each investigated method to compare the sampling quality.
This can qualitatively be compared to $\sim 2 \cdot 10^4$ samples and their weights obtained from a converged \texttt{VEGAS+} integrator.
For fKLD, we observe that the horizontal line corresponding to the $K^*(892)$ resonance is not clearly visible compared to the results of \texttt{VEGAS+}, rKLD, and FAB. Furthermore, the crossing of the horizontal and diagonal lines appears as a blurred region for fKLD.
This observation is consistent with figure~\ref{fig:lambdac_marginal_hist}, where we provide marginal histograms of the invariant mass distribution for $m_{pK^-}^2$ and $m_{K^-\pi^+}^2$. 
We average the results for three models trained with different random seeds and visualize the standard deviation per histogram bin.
Overall, the binned samples obtained from the normalizing flows trained with fKLD show the largest standard deviations from samples obtained with rejection sampling.
Flows trained with rKLD and FAB produce histograms that exhibit all features of the ME structure and provide the best results.

\begin{figure}[t]
  \centering
  \begin{subfigure}{0.49\textwidth}
      \includegraphics[width=\textwidth]{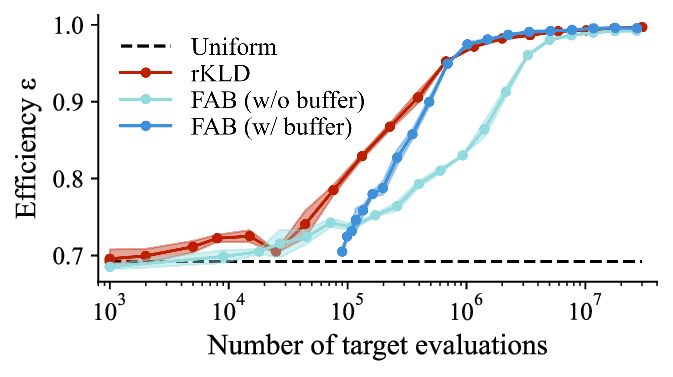}
      \caption{$\Lambda_c^+ \rightarrow pK^-\pi^+$}
      \label{fig:efficiency_target_eval_lambdac}
  \end{subfigure}
  \begin{subfigure}{0.49\textwidth}
      \includegraphics[width=\textwidth]{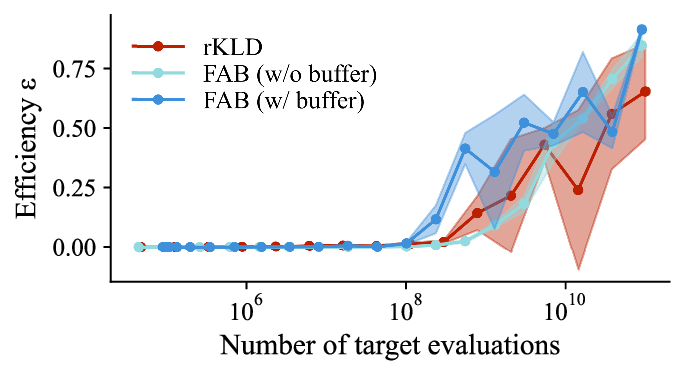}
      \caption{$e^+e^- \rightarrow t \bar{t}, t \rightarrow W^+ b, \bar{t} \rightarrow W^- \bar{b}.$}
      \label{fig:efficiency_target_eval_ee_ttbar}
      %\begin{justify}
       % \scriptsize
        %$~^{*}$\textbf{Comment:}\quad
        %Results for FAB (w/ buffer) are only averaged over two random seeds as one run diverged during training.
    %\end{justify}
  \end{subfigure}
  \caption{Importance sampling efficiency depending on the number of target evaluations required during training.}
  \label{fig:efficiency_target_eval}
\end{figure}

When training with the same number of target evaluations, rKLD and FAB with the replay buffer provide the best results with the highest importance sampling efficiency and a distribution that is most similar to the test data set (see table~\ref{table:performance_metrics}). 
Additionally, the integral estimate computed with samples from the flow has the lowest standard deviations.
The change in the sampling efficiency when increasing the number of target evaluations is shown in figure~\ref{fig:efficiency_target_eval_lambdac}. We can observe that $\epsilon$ improves for rKLD more rapidly than for FAB w/o buffer. 
All normalizing flow models show the same importance sampling efficiency of approximately $70 \%$ for a low number of target evaluations. 
This corresponds to the efficiency obtained from rejection sampling with a uniform proposal distribution (shown as a black, dashed line in figure \ref{fig:efficiency_target_eval_lambdac}).
This behavior can be explained by the fact that the normalizing flows have a uniform base distribution and are initialized with an identity mapping, resulting in a uniform distribution at the beginning of training. 
For FAB w/ buffer, samples have to be generated before the start of training to fill the replay buffer, which results in an offset in the number of target evaluations.
During training, the normalizing flow parameters are updated by sampling from the prioritized replay buffer based on~$\bar{w}_\mathrm{AIS}$, which aids the model significantly and the importance sampling efficiency improves at a faster rate compared to the flows trained with rKLD.
Since FAB evaluates the target several times in each iteration dependent on the chosen number of HMC steps and intermediate AIS distributions, it can benefit from additional information about the structure of the ME.
Overall, flows trained with rKLD and both FAB approaches have no problem in modeling the complex 2D distribution.

\begin{table*}[t]
    \caption{
        Overview of performance metrics for the different MEs and methods.
    }
    \label{table:performance_metrics}
    \centering
    \scriptsize
    \addtolength{\tabcolsep}{-0.3em}
    \begin{tabularx}{\linewidth}{
        X
        !{\color{white}\ }S[table-format=1.4(1.4)]
        !{\color{white}\ }S[table-format=2.2(1.2)]
        !{\color{white}\ }S[table-format=4.0(1.0)]
        c
        !{\color{white}\ }S[table-format=1.3(1.3)]
        !{\color{white}\ }S[table-format=2.2(2.2)]
        !{\color{white}\ }S[table-format=4.1(2.1)]
    }
        \toprule
        & 
            \multicolumn{3}{c}{$\Lambda_c^+ \rightarrow pK^-\pi^+$} &
            & 
            \multicolumn{3}{c}{$e^+e^- \rightarrow t\bar{t}, t \rightarrow W^+ b, \bar{t} \rightarrow W^- \bar{b}$} \\

        \cmidrule{2-4}
        \cmidrule{6-8}
        & \multicolumn{1}{c}{$\mathbb{E}_p \left[ \log(p/q_\theta)\right] \, \downarrow$} & 
            \multicolumn{1}{c}{$\epsilon (\unit{\percent}) \, \uparrow$} & 
            \multicolumn{1}{c}{$\bar{I}$} & 
            & 
            \multicolumn{1}{c}{$\mathbb{E}_p \left[ \log(p/q_\theta)\right] \, \downarrow$} & 
            \multicolumn{1}{c}{$\epsilon (\unit{\percent}) \, \uparrow$} & 
            \multicolumn{1}{c}{$\bar{I}$} \\
        \midrule
        \texttt{VEGAS+} &
            \textemdash &
            67.52(0.21) &
            8926(2) &
            &
            \textemdash &
            0.02(0.01) &
            1761(309) \\ \addlinespace[.25em]
        fKLD & 
            9.1581(11) & 
            87.02(8) & 
            8925(3) & 
            &
            8.35(16) & 
            1.75(1.26) & 
            2116(9) \\ \addlinespace[.25em]
        rKLD & 
            \cellcolor{accblue!25} 9.0978(5) & 
            \cellcolor{accblue!25} 99.67(1) & 
            8924(4) &
            & 
            \cellcolor{accblue!25} 7.74(0.02) & 
            56.51(40.14) & 
            2267(88) \\ \addlinespace[.25em]
        FAB (w/o buffer) & 
            9.1009(3) & 
            99.26(8) & 
            8912(7) &
            & 
            7.79(3) & 
            84.25(4.51) & 
            2208(1) \\ \addlinespace[.25em]
        FAB (w/ buffer) & 
            \cellcolor{accblue!25} 9.0988(5) & 
            \cellcolor{accblue!25} 99.56(5) & 
            8911(2) &
            & 
            \cellcolor{accblue!25} 7.747(2) & 
            \cellcolor{accblue!25} 90.59(1) & 
            2207(0.1) \\
        \bottomrule
    \end{tabularx}
\end{table*}

%\footnotetext[1]{\textbf{Note:} Values are averaged over two random seeds for FAB w/ buffer, as one run diverged during training.}

%\vspace{-0.15cm}
\paragraph{$e^+e^- \rightarrow t\bar{t}, t\rightarrow W^+ b, \bar{t} \rightarrow W^- \bar{b}$.} 

\begin{figure*}[t]
    \centering
    \includegraphics[width=\textwidth]{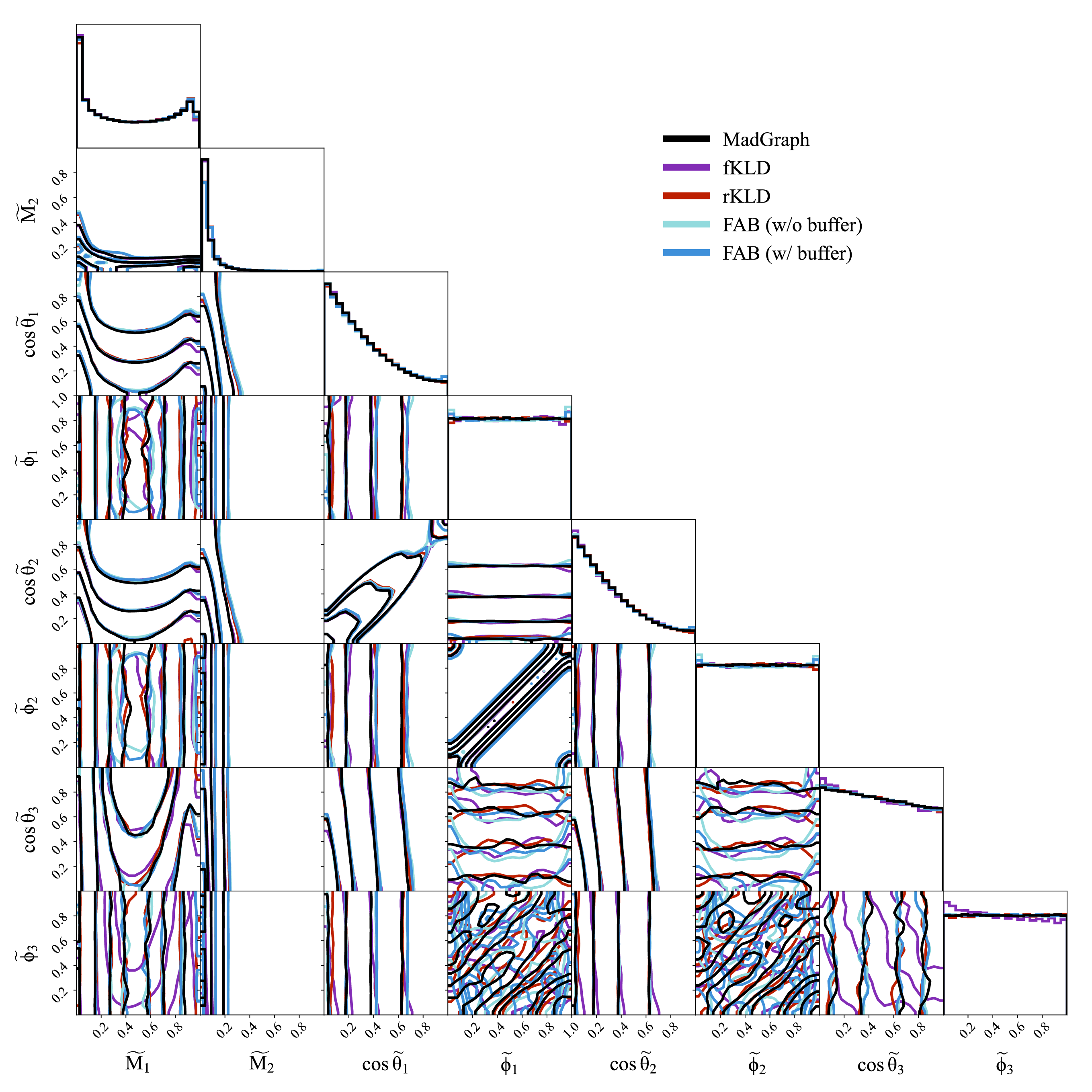}
    \caption{Corner plot with samples from \texttt{MadGraph} and from the best normalizing flows for each method.}
    \label{fig:ee_ttbar_corner_plot}
\end{figure*}

We compare the corner plots of $10^6$ flow samples for each method to samples from the training data set generated with \texttt{MadGraph} in figure~\ref{fig:ee_ttbar_corner_plot}. 
The latter serve as a ground truth and illustrate challenging properties of MEs: 
Peaks at the boundary and correlations between dimensions are difficult for normalizing flows. 
Here, we visualize samples from the flow directly on the unit hypercube and one would need to apply the inverse \texttt{RAMBO} transformation~\citep{plaetzer:2013} to obtain physical information for each outgoing particle. 
We observe that the normalizing flows especially deviate from the \texttt{MadGraph} distribution close to the phase space boundary which can be observed for all investigated methods. 
Correlations between dimensions, for example between the angles $\widetilde{\mathrm{\phi}}_1 - \widetilde{\mathrm{\phi}}_3$ as well as $\widetilde{\mathrm{\phi}}_2 - \widetilde{\mathrm{\phi}}_3$, appear to be challenging.
While we report the performance metrics based on $\sim 10^5$ samples from an optimized \texttt{VEGAS+} integrator in table~\ref{table:performance_metrics}, we do not include the results in the corner plot since we observe large deviations explained by low efficiencies.
When considering the number of target evaluations in figure~\ref{fig:efficiency_target_eval_ee_ttbar}, the flow benefits from the AIS procedure and the sampling of batches from the buffer dependent on the AIS importance weights. 
Samples from regions where the flow is a poor approximation of the target have a high weight and are predominantly used in gradient updates, resulting in a significantly higher importance sampling efficiency with fewer target evaluations. 
Therefore, FAB w/ buffer reaches an efficiency of approximately \SI{40}{\percent} with an order of magnitude fewer samples than rKLD and FAB without the buffer. 
We observe that one of the three training runs of FAB w/ buffer diverged and exclude it from figure~\ref{fig:efficiency_target_eval_ee_ttbar} and the performance evaluation in table~\ref{table:performance_metrics}.

\section{Summary and Conclusion}
We have transferred FAB~\citep{midgley:2023}, which utilizes AIS with HMC as a transition operator, from molecular configuration modeling to HEP, building on recently introduced, differentiable implementations of matrix elements. 
We have demonstrated that training FAB with a prioritized replay buffer is a promising approach for improving the efficiency for event generation, since passing the flow samples through an AIS chain guides training at early stages.
In the future, we plan to scale this approach to more complex particle-interaction processes and assess the performance improvement in greater detail. 
Including information about the individual matrix element contributions via multi-channeling will likely lead to further performance improvements~\citep{bothmann:2020, heimel:2023, heimel:2024}.
Additionally, we plan to extend this conceptual work with differentiable implementations of the parton density~\citep{carrazza:2021, heimel:2024_diffable} to be applicable to proton-proton collisions.
Overall, this work has the potential to improve the quality and speed of sampling methods employed at the LHC and facilitate the efficient analysis of high-luminosity events.
\looseness=-1

\paragraph{Code and Data Availability.}
The code is available on \href{https://github.com/annalena-k/FAB-meets-diffME}{GitHub} and the data can be downloaded from \href{https://doi.org/10.17617/3.UZ786R}{Edmond}~\citep{kofler_data:2024}.
%\looseness=-1

%%%%%%%%%%% Acknowledgements %%%%%%%%%%%
\begin{ack}
The authors thank the reviewers of MLST and the NeurIPS workshop \textit{Machine Learning and the Physical Sciences, 2024} for their helpful suggestions and interesting questions which helped to improve the manuscript.
AK thanks Timothy Gebhard for extensive support in reviewing and formatting the draft as well as Ludwig Burger and Nicole Hartman for proof-reading and correcting the manuscript. 
The authors thank Matthew Feickert for help with running \texttt{MadGraph}. 
AK thanks Peter Lepage for answering questions about running~\texttt{VEGAS+}.
The computational work described in this manuscript was performed on the RAVEN cluster of the Max Planck Computing and Data Facility (MPCDF).
We thank the International Max Planck Research School for Intelligent Systems (IMPRS-IS) for supporting AK. 
MK is supported by the US Department of Energy (DOE) under grant DE-AC02-76SF00515. LH is supported by the Excellence Cluster ORIGINS, which is funded by the Deutsche Forschungsgemeinschaft (DFG, German Research Foundation) under Germany’s Excellence Strategy - EXC-2094-390783311.
We also thank the Munich Institute for Astro-, Particle and BioPhysics (MIAPbP) which is funded by the Deutsche Forschungsgemeinschaft (DFG, German Research Foundation) under Germany's Excellence Strategy - EXC-2094 - 390783311, as this work was partially performed at the MIAPbP workshop on Differentiable and Probabilistic Programming for Fundamental Physics.

\textit{Used software:} This work has made use of many open-source Python packages, including 
\texttt{blackjax}~\citep{blackjax:2024}, 
\texttt{ComPWA}~\citep{compwa:2023},
\texttt{corner}~\citep{corner:2016},
\texttt{distrax}~\citep{deepmind_jaxecosys_distrax:2020}, \texttt{haiku}~\citep{haiku:2020}, 
\texttt{JAX}~\citep{jax:2018}, 
\texttt{FAB-JAX}~\citep{midgley:2023}, 
\texttt{MadJAX}~\citep{heinrich:2022},
\texttt{MadGraph}~\citep{madgraph:2014},
\texttt{matplotlib}~\citep{matplotlib:2007},
\texttt{numpy}~\citep{numpy:2020}, 
\texttt{pylhe}~\citep{pylhe:2024}
%\texttt{PYVEGAS}~\citep{lepage:1978},
\texttt{SymPy}~\citep{sympy:2017}, 
\texttt{TensorWaves}~\citep{tensorwaves}, and
\texttt{VEGAS+}~\citep{lepage:1978, lepage:2021}.
The accessible color schemes in our figures are based on~\citep{accessible_color_scheme:2021}.

\end{ack}

%%%%%%%%%%%  APPENDIX  %%%%%%%%%%%%%
\newpage 
\appendix
\section*{Appendix}

\section{Transformation of Dalitz plot to unit hypercube}
\label{sec:trafo_DP_to_unit_hypercube}

\begin{figure*}[ht]
    \centering
    \includegraphics[width=0.8\textwidth]{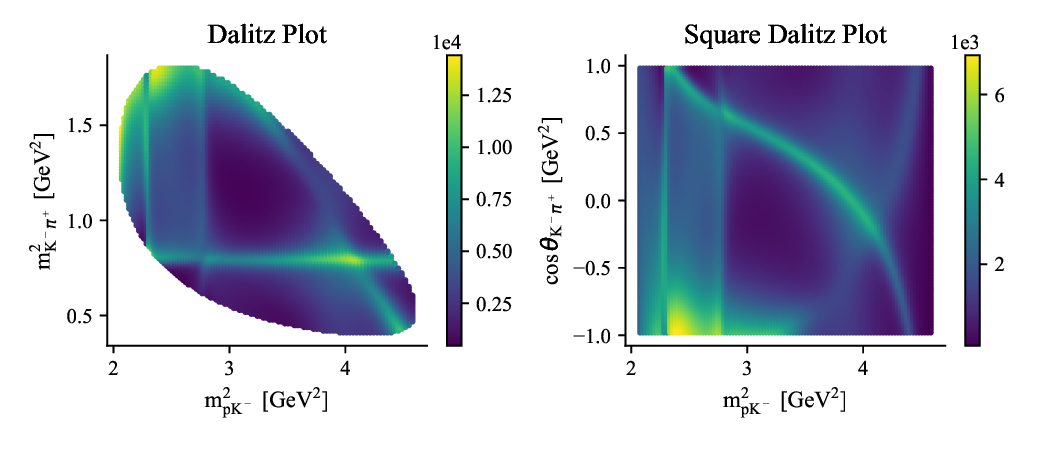}
    \caption{Dalitz plot of $\Lambda_c^+ \rightarrow pK^- \pi^+$ decay and transformation to square Dalitz plot.}
    \label{fig:lambdac_square_dalitz_plot}
\end{figure*}

While a Dalitz plot is the ideal visualization of the outgoing particles of a three-particle decay from a physics perspective, it is not ideal for normalizing flow training since the flow has to learn the support of the distribution.
To prevent the normalizing flows from assigning non-zero probability density outside the phase space boundary, we map the Dalitz plot coordinates $(m_{pK^-}^2, m_{K^-\pi^+}^2)$ to the unit hypercube with a differentiable and invertible transformation.
First, we convert the Dalitz plot into the square Dalitz plot by transforming one of the invariant masses, e.g., $m_{K^- \pi^+}^2$ to the helicity polar angle $\cos \theta_{K^- \pi^+}$ \cite{byckling:1973}. %[p.105ff.]
The mapping between the two variables is defined as
\begin{equation*}
    \renewcommand{\arraystretch}{1.5} % Increase row height by 1.5x
    \begin{array}{l}
    m_{K^- \pi^+}^2 = m_{\pi^+}^2 + m_{K^-}^2 \\
    \hspace{1.7cm} + ~\frac{1}{2 m_{pK^-}^2} \left(m_{\Lambda_c^+}^2 - m_{pK^-}^2 - m_{\pi^+}^2 \right) \left( m_{pK^-}^2 + m_{K^-}^2 - m_{p}^2 \right)\\
    \hspace{1.7cm} - ~\frac{1}{2 m_{pK^-}^2 } \cos \theta_{K^-\pi^+} \cdot \lambda^{\frac{1}{2}} (m_{\Lambda_c^+}^2, m_{pK^-}^2, m_{\pi^+}^2) \cdot \lambda^{\frac{1}{2}} (m_{pK^-}^2, m_{K^-}^2, m_{p}^2) ~.
	% from Byckling equation V 1.8
    % 0 = Lambda_c^+, 1 = \pi^+, 2 = K^-, 3 = p
    \end{array}
\end{equation*}
where the masses $m_{\pi^+}, m_{K^-},$ and $m_p$ are constants and the kinematic K\"{a}ll\'{e}n function is defined as
\begin{equation*}
	\lambda(x,y,z) = x^2 + y^2 + z^2 - 2xy - 2xz - 2yz~.
\end{equation*}
After applying the non-linear transformation from $(m_{pK^-}^2, m_{K^-\pi^+}^2)$ to $(m_{pK^-}^2, \cos \theta_{K^-\pi^+})$, we can visualize the resulting distribution in a square Dalitz plot in figure~\ref{fig:lambdac_square_dalitz_plot}.
We can observe that the previously horizontal line gets mapped to an almost diagonal line.

\section{Hyperparameters} \label{sec:hyperparameters}
\paragraph{Data generation and training.}
The training data for fKLD consists of $10^4$ samples for \mbox{$\Lambda_c^+ \rightarrow pK^-\pi^+$} generated with rejection sampling and of $10^6$ samples for \mbox{$e^+e^- \rightarrow t\bar{t}$}, \mbox{$t \rightarrow W^+b$}, \mbox{$\bar{t} \rightarrow W^-\bar{b}$} produced with \texttt{MadGraph}~\citep{madgraph:2014} with a center of mass energy of \SI{1}{\tera\electronvolt}. 
The test data sets are generated equivalently.
The dimensionality of the resulting ME is defined by the number of degrees of freedom as $3n - 4$ which we explain for illustration for the eight-dimensional case: 
Each of the four outgoing particles is defined by its energy and three-dimensional momentum vector, resulting in 16 overall degrees of freedom. 
Conservation of energy and momentum reduces this number by four dimensions. 
Additionally, we know the masses of each stable outgoing particle, resulting in a $12 - 4 = 8$ dimensional phase space.
For fKLD, the normalizing flows are trained with these datasets for 20 epochs in 2D and for 200 epochs in 8D. 
For rKLD, we train for $3 \times 10^4$ iterations (2D) and for $10^8$ iterations (8D). The values for FAB were chosen based on the FAB hyperparameters such that we perform the same number of target evaluations during training compared to rKLD, resulting in $3 \times 10^3$ iterations (2D) and $10^7$ iterations (8D). All compared models are trained on Nvidia A100 GPUs.

\paragraph{Normalizing Flow.}
We use normalizing flows based on coupling layers and rational-quadratic spline transformations~\citep{durkan:2019} implemented in \texttt{distrax}~\citep{deepmind_jaxecosys_distrax:2020} and \texttt{haiku}~\citep{haiku:2020}. For the base distribution, we choose a uniform distribution over the unit hypercube.
Since changes in the normalizing flow hyperparameters affect the expressivity of the density estimator equally for all investigated methods, we perform hyperparameter tuning only for rKLD training because it has the shortest run time. 
We only tune the most important hyperparameters related to the architecture of the normalizing flow since this is computationally costly.
For \mbox{$\Lambda_c^+ \rightarrow pK^-\pi^+$}, we subsequently vary the number of spline bins \mbox{$n_\mathrm{b} \in [4,6,8,10,12,14,16]$}, the number of flow transformations \mbox{$n_\mathrm{t} \in [4,6,8,10,12,14,16]$}, and the number of neurons per hidden layer \mbox{$n_\mathrm{n} \in [10, 50, 100, 150, 200, 250, 300]$} for the fully connected conditioner network with two layers. 
For the 8D ME, we compare loss values and efficiencies for \mbox{$n_\mathrm{b}, n_\mathrm{t} \in [6, 8, 10, 12, 14]$, and $n_\mathrm{n} \in [200, 250, 300, 350, 400]$}.
Considering the trade-off between expressivity and increase in optimization time, we select the following values based on the final training loss, validation loss, and efficiency: 
For $\Lambda_c^+ \rightarrow pK^-\pi^+$, the number of bins and transformations is $n_\mathrm{t} = n_\mathrm{b} = 10$, and a conditioner network with two hidden layers and 100 neurons each is used. For the 8D ME, we set \mbox{$n_\mathrm{b} = n_\mathrm{t} = 14$}, and use 400 neurons for each of the two hidden layers.

\paragraph{FAB.} 
Additional hyperparameters have to be selected for FAB. 
We use two (linearly spaced) intermediate distributions~$M$ in the AIS sequence with a HMC transition operator containing a single iteration and three leap frog steps. 
The initial HMC acceptance rate is set to~$p_\mathrm{acc} = 0.65$ and is tuned dependent on the number of actually accepted samples. 
Furthermore, we have to specify the number of gradient updates per iteration~$L=4$~(2D) and $L=2$~(8D) in the case of buffered training.
The most important variable in hyperparameter optimization is the HMC step size, since it depends on the support of the target distribution.
To obtain a suitable estimate, we start one FAB run with an arbitrary step size and observe how the value is adjusted during training. 
We adopt the converged values of $l_\mathrm{init} = 0.05$ for the two-dimensional and $l_\mathrm{init} = 0.005$ for the eight-dimensional ME for all subsequent runs. 
We do not optimize the size of the FAB replay buffer since we do not expect a significant influence on the performance.
We only take into account that the replay buffer should be sufficiently large such that sampling data points from the buffer is informative for normalizing flow training.
Therefore, we set the minimal size of the buffer to 10 times the batch size (i.e., $10^4$) and the maximal buffer size to 100 times the batch size (i.e., $10^5$).

\paragraph{Optimization.}
We use the \texttt{Adam}~\citep{kingma:2015} optimizer with a learning rate of $3 \times 10^{-4}$ and train with a batch size of $10^3$. We employ the gradient clipping scheme developed for FAB in all our runs, where we dynamically clip the gradient norm to 20 times the median of the last 100 gradient values and ignore very large gradients that are a factor 20 times larger than this median value~\citep{midgley:2023}.
While it is not necessary to use a scheduler in the 2D case, we employ a warm-up and cosine decay learning rate schedule for the eight-dimensional ME. 
We train with an initial and final learning rate of $10^{-5}$ as well as a peak learning rate of $3 \times 10^{-4}$ which is reached after 10 epochs (for fKLD) and after $10^3$ iterations (for rKLD and FAB).

\paragraph{Baseline.}
We compare our results with the physics-agnostic integral estimation method \texttt{VEGAS+}~\citep{lepage:1978, lepage:2021}. 
This grid-based optimization method subdivides the support into a regular, rectangular grid and estimates the integral contribution of each subspace. 
With this information, the grid is iteratively updated to focus on regions with large contributions and the value for the integral estimate is calculated as a weighted average over multiple runs.
Through a combination of stratified and importance sampling, \texttt{VEGAS+} is fast and efficient and can account for correlations between dimensions. 
For the 2D ME, we choose a \texttt{VEGAS+} grid with 64 bins in each dimension, a damping factor of $\alpha=0.5$, two warm-up iterations with $10^3$~evaluations per iteration, followed by 8~iterations with $2 \cdot 10^5$ evaluations per iteration for the integral estimates. 
For the 8D ME, we double the grid to 128 bins per dimension, keep the damping factor, and increase the number of evaluations to $10^4$ for each of the two warm-up iterations. 
The integral estimate is obtained from 8 iterations with $10^5$ evaluations.
We optimized these hyperparameters to the best of our knowledge.
To make sure that \texttt{VEGAS+} converged, we perform checks like increasing the number of evaluations by a factor of 10 per iteration which provide stable results and do not show significant deviations in the integral estimate for both examples. 
However, the low importance sampling efficiency and large deviation on the integral estimate for the 8D ME (c.f., table \ref{table:performance_metrics}) indicate that \texttt{VEGAS+} struggles to adapt to the distribution.
It is important to note that the results are obtained without multi-channeling since this physics information based on the contributing matrix elements is not provided to the normalizing flows either.
Multi-channeling would lead to a performance improvement for both \texttt{VEGAS+} as well as flow-based methods~\citep{heimel:2023, heimel:2024}.
Although the \texttt{VEGAS+} optimization is significantly faster than training a normalizing flow, the flexibility of the latter results in higher importance sampling efficiencies.
Since the main goal of training ME surrogates is to provide an optimally pre-trained model~\citep{heimel:2024}, the training time is amortized when generating a large number of events.

\section{Relationship of sampling efficiency and unweighting efficiency}\label{sec:efficiencies}

\paragraph{Importance sampling efficiency.} 
The (importance) sampling efficiency~$\epsilon$ (c.f.~\ref{eq:sampling_efficiency}) can be derived from the effective sample size~$ESS$ which allows the performance comparison of different Monte Carlo methods like Markov Chain Monte Carlo~(MCMC) or importance sampling~(IS) based on a set of weighted samples.
If we draw $N$ samples from a less-than-ideal MCMC or IS proposal distribution~$q(x)$, the $ESS$ indicates the number of independent samples that these would be equivalent to if drawn directly from the target distribution~$p(x)$.
Therefore, the effective sample size can be defined proportional to the ratio of the variance of an ideal MC~estimator (i.e.,~sampling from the target $p$) and the variance of the less-than-ideal MCMC or IS estimator (i.e.,~sampling from the proposal~$q$)~\citep{martino:2017}. Through derivations outlined in~\citep{elvira:2022}, the $ESS$ can be related to the variance of the importance weights via 
\begin{equation*}
   ESS = \frac{N}{1 + \mathrm{Var}_{q(x)}[w]}, 
\end{equation*}
and to 
the estimate
\begin{equation}
    \widehat{ESS} = N ~\frac{\left(\frac{1}{N}\sum_{i=1}^N w_i\right)^2}{\frac{1}{N}\sum_{i=1}^N w_i^2} 
    %= \frac{\left(\sum_{i=1}^N w_i\right)^2}{\sum_{i=1}^N w_i^2}
    = \frac{1}{\sum_{i=1}^N \bar{w}_i^2}~.
    \label{eq:deriv_sampling_efficiency}
\end{equation}
with the normalized importance weights \begin{equation*}
    \bar{w}_i = \frac{w_i }{\sum_{j=1}^N w_j}~.
\end{equation*}
The importance sampling efficiency~$\epsilon$ as defined in~\ref{eq:sampling_efficiency} corresponds to the normalization of~\ref{eq:deriv_sampling_efficiency}.

\paragraph{Unweighting efficiency.} 
The unweighting efficiency stems from the approach of refining samples obtained from the less-than-ideal proposal distribution~$q(x)$ by keeping only a fraction of the samples in proportion to the ratio of the target distribution~$p(x)$ and the proposal~$q(x)$~\citep{klimek:2020}.
The so-called \textit{raw weight} $w_i = \frac{p(x_i)}{q(x_i)}$ corresponds to the importance weight in the MCMC and IS setting.
The definition of the unweighting efficiency \citep{klimek:2020, bothmann:2020, danziger:2022} as 
\begin{equation}
    \epsilon_\mathrm{uw} = \frac{\frac{1}{N} \sum_{i=1}^N w_i} {w_\mathrm{max}}
\end{equation}
can be motivated in the following: 
In regions where the proposal distribution $q$ overestimates the target (i.e.,~$w_i < 1$), only a fraction of the original samples proportional to~$w_i$ should be retained. 
However, in regions where the proposal underestimates the target, it is not possible to generate additional samples to match the target's density. 
Therefore, all available samples in those regions are kept which has to be compensated by reducing the retained fraction in overestimated regions. 
Such an adjustment maintains the correct relative shape of the distribution~\citep{klimek:2020}.
This explanation is equivalent to applying rejection sampling where the probability to keep or reject a sample is defined as the raw weight normalized by the (pre-computed) maximal weight in the integration volume~$w_\mathrm{rel} = w_i / w_\mathrm{max}$. 
A~sample is retained if a uniformly sampled random number~$R$ is smaller than~$w_\mathrm{rel}$~\citep{backes:2021, danziger:2022}.
Finally, the unweighting efficiency can be computed as the average raw weights rescaled by $w_\mathrm{max}$~\citep{bothmann:2020, danziger:2022}.

\paragraph{Relationship of efficiencies.} 
In \citep{martino:2017}, different formulations of the generalized effective sample size are explored based on a set of required and desirable conditions. 
They conclude that defining the $ESS$ as \mbox{$1 / \sum_{i=1}^N \bar{w}_i^2$} (related to~$\epsilon$) and \mbox{$1 / \max(\bar{w}_1, ..., \bar{w}_N)$} (related to~$\epsilon_\mathrm{uw}$) are proper and stable formulations.
While both efficiencies suffer from large outlier weights, the unweighting efficiency is directly affected through $w_\mathrm{max}$ in the denominator as reported in \citep{bothmann:2020, gao:2020_evgen_nfs}. 
For this reason, we choose to report the performance based on the importance sampling efficiency in the main body of this work and provide additional estimates of the unweighting efficiency in table \ref{table:efficiencies}. 
We do not apply bootstrap techniques that are designed to mitigate large outliers in the weight distribution~\citep{gao:2020_evgen_nfs}.

\begin{table*}[t]
    \caption{
        Comparison of importance sampling and unweighting efficiency for the different MEs and methods.
    }
    \label{table:efficiencies}
    \centering
    \scriptsize
    \addtolength{\tabcolsep}{0.5em}
    \begin{tabularx}{0.9\textwidth}{
        l
        !{\color{white}\ }S[table-format=2.2(1.2)]
        !{\color{white}\ }S[table-format=2.2(1.2)]
        c
        !{\color{white}\ }S[table-format=2.2(2.2)]
        !{\color{white}\ }S[table-format=3.2(2.2)]
    }
        \toprule
        & 
        \multicolumn{2}{c}{$\Lambda_c^+ \rightarrow pK^-\pi^+$} &
        & 
        \multicolumn{2}{c}{$e^+e^- \rightarrow t\bar{t}, t \rightarrow W^+ b, \bar{t} \rightarrow W^- \bar{b}$} \\

        \cmidrule{2-3}
        \cmidrule{5-6}
        & \multicolumn{1}{c}{$\epsilon (\unit{\percent}) \, \uparrow$} & 
            \multicolumn{1}{c}{$\epsilon_\mathrm{uw} (\unit{\percent}) \, \uparrow$} & 
            & 
            \multicolumn{1}{c}{$\epsilon (\unit{\percent}) \, \uparrow$} & 
            \multicolumn{1}{c}{$\epsilon_\mathrm{uw} (\unit{\percent}) \, \uparrow$} \\
        \midrule
        \texttt{VEGAS+} &
            67.52(21) &
            16.30(42) &
            &
            0.02(1) &
            < 0.01(0) \\ \addlinespace[.25em]
        fKLD & 
            87.02(8) & 
            22.50(37) & 
            &
            1.75(1.26) & 
            0.02(1) \\ \addlinespace[.25em]
        rKLD & 
            \cellcolor{accblue!25} 99.67(1) & 
            \cellcolor{accblue!25} 75.42(1.87) &
            & 
            56.52(40.14) & 
            0.29(20) \\ \addlinespace[.25em]
        FAB (w/o buffer) & 
            99.26(8) & 
            67.59(1.38) &
            & 
            84.25(4.51) & 
            \cellcolor{accblue!25} 1.15(27) \\ \addlinespace[.25em]
        FAB (w/ buffer) & 
            \cellcolor{accblue!25} 99.56(5) & 
            \cellcolor{accblue!25} 68.69(89) &
            & 
            \cellcolor{accblue!25} 90.59(1) & 
            0.67(7)\\
        \bottomrule
    \end{tabularx}
\end{table*}
 
\medskip

%%%%%%%%%%%  BIBLIOGRAPHY  %%%%%%%%%%%%%

\bibliographystyle{iopart-num}
\bibliography{literature}

\end{document}